\documentclass[useAMS, usenatbib, usegraphicx]{mn2e}
\usepackage{times}

\newcommand{\ltsimeq}{\raisebox{-0.6ex}{$\,\stackrel
           {\raisebox{-.2ex}{$\textstyle <$}}{\sim}\,$}} 

\title[Resumption of mass accretion in RS~Oph]{Resumption of mass accretion in RS~Oph}

\author[H. L. Worters et al.]{H. L. Worters$^1$,
S. P. S. Eyres$^1$,
G. E. Bromage$^1$,
J. P. Osborne$^2$,\\  
$^1$Centre for Astrophysics, University of Central Lancashire,
Preston, PR1 2HE, UK\\
$^2$Department of Physics and Astronomy, University of Leicester,
Leicester, LE1 7RH, UK}

\date{Received; in original form}

\begin{document}
\label{firstpage}
\maketitle
\begin{abstract}
The latest outburst of the recurrent nova RS~Oph occurred in 2006
February.  Photometric data presented here show evidence of the
resumption of optical flickering, indicating re-establishment of
accretion by day~241 of the outburst.  Magnitude variations of up to
0.32~mag in V-band and 0.14~mag in B on time-scales of 600-7000~s are
detected.  Over the two week observational period we also detect a
0.5~mag decline in the mean brightness, from $\rm V \approx 11.4$ to
$\rm V \approx 11.9$, and record $\rm B \approx 12.9$~mag.  Limits on
the mass accretion rate of $\sim 10^{-10} \leq \dot{M}_{\rm acc} \leq
10^{-9} M_{\sun} \rm yr^{-1}$ are calculated, which span the range of
accretion rates modeled for direct wind accretion and Roche lobe
overflow mechanisms.  The current accretion rates make it difficult
for thermonuclear runaway models to explain the observed recurrence
interval, and this implies average accretion rates are typically
higher than seen immediately post--outburst.
\end{abstract}

\begin{keywords}
stars: individual: RS~Oph -- novae, cataclysmic variables -- mass-loss
stars: winds, outflows
(stars:) binaries: symbiotic
\end{keywords}

\section{Introduction}
\label{sec-intro}

Recurrent novae (RNe) are interacting binary systems in which multiple
nova outbursts have been observed.  Both thermonuclear runaway and
accretion models have been hypothesised as the outburst mechanism in
these systems \citep{Ken86}.  While thermonuclear is generally the preferred
mechanism, there are problems with the high accretion rate required
given the short outburst recurrence interval. The recurrent nova
RS~Ophiuchi has undergone six recorded outbursts in the last 108 years
\citep{Opp93}, the most recent occurring on 2006 February 12, which we
take as day 0 \citep{Hir06}.  RS~Oph consists of a white dwarf primary
accreting material from a red giant secondary within a nebula formed
from the red giant wind.  Attempts to classify the secondary component
have resulted in suggestions ranging from K0 III \citep{Wal69} to M4
III \citep{Boh89}, with several concluding M2 III to be most likely
\citep{Bar69,Ros82,Bru86,Opp93}. The white dwarf in the system is
close to the Chandrasekhar mass limit \citep{Dob94}, hence the ratio
of mass accreted to mass ejected will determine whether RS~Oph is a
potential supernova Ia progenitor \citep{Sok06}.

The quiescent characteristics of RS~Oph have led to its classification
as a symbiotic star, although with a weak hot-component spectrum.
Most symbiotic stars do not exhibit the variability on time-scales of
minutes seen in cataclysmic variables \citep{Sok01}, yet
short-time-scale, aperiodic variations in optical brightness have long
been known in RS~Oph in its quiescent state \citep{Bru86}.  These
stochastic or aperiodic brightness variations are known as flickering,
with `strong' flickering being of the order of a few tenths of
magnitudes (Sokoloski et al. 2001).  While symbiotic stars are a
heterogenous class, other members show similarities to RS~Oph that are
applicable here.

To date, there have been no reported observations of the
re-establishment of optical flickering in the immediate post-outburst
phase of a recurrent nova, a fact that contributes to our uncertainty
of the nature of the outburst mechanism.  Observations by
\citet{Zam06} on day 117 (2006 June 9) show no flickering of amplitude
above 0.03~mag in B, from which they conclude that an accretion disc
around the white dwarf has been destroyed as a result of the 2006
outburst.  The lightcurve reached a post--outburst minimum in 2006
September.  Following discovery of rebrightening \citep{Bod06a}, we
monitored RS~Oph photometrically for two weeks in B and V bands,
detecting the resumption of optical flickering \citep{Wor06}.

\section{Observations}
\label{sec-observations}

Observations of duration 37 to 118~minutes were made on eleven nights
between 2006 October 11 and 24, the shorter observations being
curtailed by cloud.  Observations were made with the South African
Astronomical Observatory (SAAO) 1-m telescope and the SAAO CCD camera,
a 1024$\times$1024 pixel SITe back-illuminated chip.  The field of
view is 5\arcmin$\times$5\arcmin, which is sufficient to include
several comparison stars close to the target, including
USNO-B1.0~0833-0368817 and -0368883.  Integration times were typically
10~s in Johnson~V (20~s in Johnson~B), with a readout time of 19~s,
allowing continuous V-band monitoring with a temporal resolution of
$\sim$30~s.  Longer exposure times were occasionally used to
compensate for poorer sky conditions.  Details of each night's
observations are given in Table 1. The 3 nights lacking data were lost
due to cloud.

Preliminary data reduction was performed using standard procedures in
IRAF.  The resulting images were then processed using CCD tasks in the
SAAO STAR package \citep[described in][]{Bal95,Cra00} to determine
aperture magnitudes of the target and selected comparison stars.

\section{Results}
\label{sec-results}

\begin{figure}
    \includegraphics[width=0.99\columnwidth]{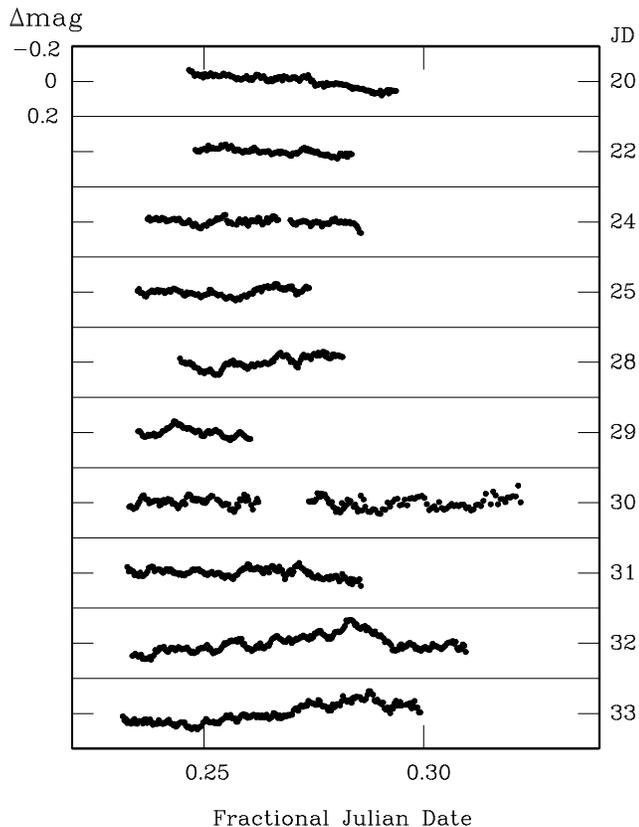}
    \caption{Ten nights' differential V-band lightcurves of RS~Oph.  Magnitudes are normalised to the mean value for each night to illustrate relative flickering amplitudes.  Numbers down the right-hand margin are $\rm JD - 2454000$.  The break in data points on the night labelled $JD=30$ is due to cloud.}
    \label{fig-rsnorm}
  \end{figure}

Figure~\ref{fig-rsnorm} shows the diversity of flickering amplitude
and time-scale present in the V-band lightcurves obtained on ten nights
of the 2~week period of observations.  Visual inspection reveals an
increase in flickering amplitude during nights towards the end of the
run.

\begin{figure}
    \includegraphics[width=0.99\columnwidth]{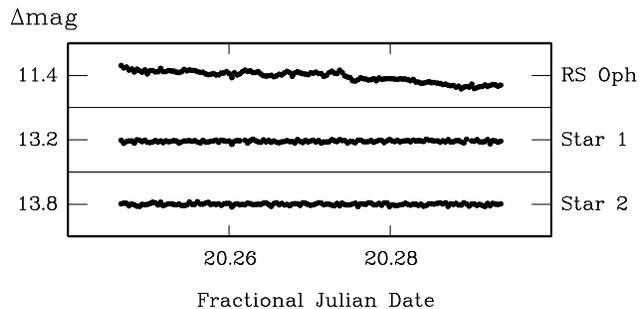}
    \caption{V-band lightcurves of RS~Oph and 2 comparison stars from 2006 Oct 11, when the weakest flickering was detected.  The ordinate for each plot spans 0.4~mag.  Amplitude variability of 0.06 mag is evident in the target.} 
    \label{fig-11diff}
  \end{figure}

\begin{figure}
    \includegraphics[width=0.99\columnwidth]{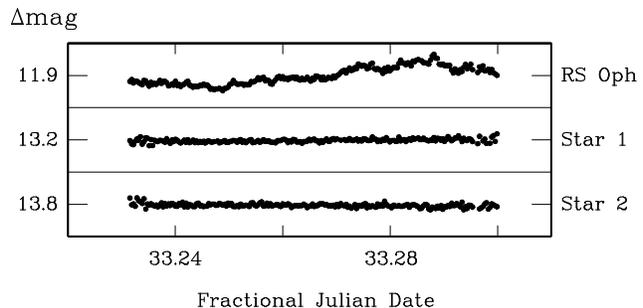}
    \caption{Example V-band lightcurves of RS~Oph and 2 comparison stars in the field.  These data are taken from 2006 Oct 24, one of the nights showing the strongest flickering observed, with amplitude 0.31 mag.  Again, the ordinate spans 0.4~mag for each plot.} 
    \label{fig-24diff}
  \end{figure}

\begin{figure}
    \includegraphics[width=0.99\columnwidth]{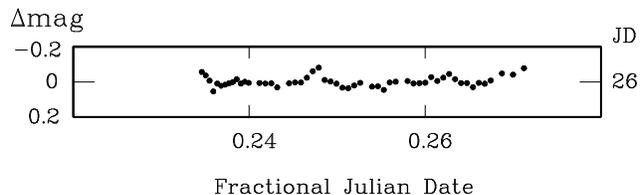}
    \caption{Differential B-band lightcurve of RSOph from 2006 Oct 17, normalised to the mean magnitude over the night.}
    \label{fig-rsnB}
  \end{figure}

Figures~\ref{fig-11diff} and ~\ref{fig-24diff} show differential
lightcurves of RS~Oph compared with two comparison stars in the field
for the nights during which we detect some of the smallest and
greatest flickering amplitudes, respectively.  Comparing the weakest
flickering detected in the target (Figure~\ref{fig-11diff}) with
brightness variations in the constant comparison stars verifies the
intrinsic variability of RS~Oph.  Flickering is also detected in the
B-band data, plotted in Figure~\ref{fig-rsnB}.

\citet{Gro06} observed a selection of symbiotic stars, performing a
statistical evaluation of the significance of flickering in the data.
They calculate mean magnitudes and standard deviations in their
variable targets ($\sigma_{\rm var}$) and comparison stars
($\sigma_{\rm comp}$).  Since the comparison stars in the field are
all $\geq 2$~mag fainter than RS~Oph, standard deviations on the value
expected for a constant star of the same brightness as the target
($\sigma '_{\rm comp}$) are derived from an empirical formula.  With
the number of counts in the data presented here being significantly
lower than the \citet{Gro06} values (a few 1000s, cf. $10^{5}$), this
method proved less reliable when applied to our data.  Two alternative
methods of deriving $\sigma ' _{\rm comp}$ were used in the current
analysis: (a) fitting a power law to the mean magnitude and
$\sigma_{\rm comp}$ values for the comparison stars, obtaining an
estimate of $\sigma ' _{\rm comp}$ in RS~Oph by extrapolation, and (b)
estimating $\sigma ' _{comp}$ by equating it to $\sigma _{comp}$ for
the brightest comparison star (13.2 mag), thus yielding very
conservative values.  All results presented here were obtained using
(b), the more conservative technique, i.e. giving larger error bars.

The ratio $R_{\rm var} = \sigma_{\rm var} / \sigma'_{comp}$ can be
used to assess the significance of the flickering.  The criteria
specified by \citet{Gro06} to determine the existence of flickering
are:\\ 
\begin{itemize}
\item $1.5 \le R_{\rm var} < 2.5$ - flickering ``probably present'', and \\ 
\item $2.5 \le R_{\rm var}$ - flickering ``definitely present''. \\
\end{itemize}

Evaluating the full dataset for each night according to the above
criteria suggests that flickering is definitely evident in RS~Oph on
all but 3 nights observed.  Even using the conservatively large error
estimates, $R_{\rm var}$ is close to the cut-off value for definite
flickering on these 3 nights.  Applying these criteria to ten minute
periods within each night's data, we detect at least probable
flickering for all ten minute periods on 6 nights, and definite
flickering for at least half of all ten minute periods on 6 nights.
Again, despite being conservative estimates, these values are very
close to $R_{\rm var}$ for definite flickering.  Table~1 shows the
mean ratio $R_{\rm var}$ averaged over all ten minute intervals for
each night, and also for each full night's data.  The statistical
analysis presented here is adequate to demonstrate that significant
flickering is detected on time-scales of ten minutes to 2 hours.

Using Equation (3) of \citet{Gro06}, we obtain V-band flickering
amplitudes ($A$) in RS~Oph ranging from 0.06~mag to 0.32~mag.  Table~1
lists flickering amplitudes derived from both the full data set for
each night, as well as mean values for ten minute intervals within
each night's data.

\begin{table*}
\begin{tabular}{lcccclccccc}
Date & JD & Day of   & Filter & $T$ & $t_{\rm exp}$& $\bar{V}$ & $\bar{R}_{\rm var}$& $R_{\rm var}$& $\bar{A}$ (mag) & $A$ (mag) \\    

 &(mid-obs)   & outburst & (Johnson)& (min) & (s) & (mag) & ($\tau=10$ min)  & ($\tau=T$) & ($\tau=10$ min)  & ($\tau=T$) \\ 
\hline	       		     
20061011 &2454020.24 &  241  &V & 68  & 10        &  11.40 & 2.38 [0.91]& 1.99  & 0.06 & 0.06 \\
20061013 &2454022.27 &  243  &V & 51  & 10        &  11.52 & 2.37 [1.08]& 3.21  & 0.06 & 0.10 \\
20061015 &2454024.26 &  245  &V & 50  & 10        &  11.50 & 1.83 [0.55]& 2.00  & 0.07 & 0.09 \\
20061016 &2454025.26 &  246  &V & 56  & 10        &  11.56 & 3.37 [0.83]& 5.56  & 0.07 & 0.12 \\
20061017 &2454026.25 &  247  &B & 52  & 20,40,90  &  12.86 & 3.30 [1.39]& 2.40  & 0.14 & 0.14 \\
20061019 &2454028.27 &  249  &V & 73  & 10        &  11.65 & 2.93 [1.64]& 4.68  & 0.10 & 0.20 \\
20061020 &2454029.25 &  250  &V & 37  & 10        &  11.67 & 2.40 [0.67]& 4.30  & 0.07 & 0.14 \\
20061021 &2454030.28 &  251  &V & 118 & 10,30     &  11.64 & 3.49 [2.90]& 4.34  & 0.21 & 0.32 \\
20061022 &2454031.26 &  252  &V & 77  & 10        &  11.81 & 2.91 [0.91]& 4.00  & 0.09 & 0.14 \\
20061023 &2454032.27 &  253  &V & 109 & 10        &  11.84 & 4.54 [1.91]& 12.02 & 0.10 & 0.29 \\
20061024 &2454033.26 &  254  &V & 100 & 10        &  11.88 & 2.51 [1.20]& 5.50  & 0.08 & 0.31 \\
\end{tabular}
\label{tab-obs}
\caption{
Observations made using the SAAO 1-m telescope and SAAO CCD. $T$ is the total duration of each night's observations, $t_{\rm exp}$ is the exposure time.  $\tau$ is the time-scale over which flickering significances ($R_{\rm var}$) and amplitudes ($A$) are calculated.  Values enclosed in square brackets are standard deviations of $\bar{R}_{\rm var}$.}
\end{table*}

\begin{figure}
    \includegraphics[width=0.99\columnwidth]{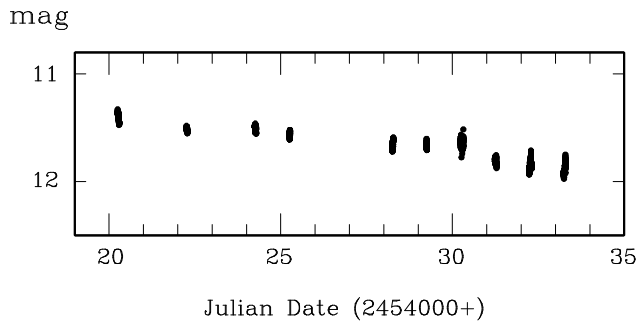}
    \caption{Decline in V magnitude of RS~Oph plotted for each night over the 2 week observational period.}
    \label{fig-rsmag}
  \end{figure}

A decrease in the mean magnitude of RS~Oph over the 2 week period is
depicted in figure~\ref{fig-rsmag}, from which the range in V
magnitude detected each night is also apparent.  The mean magnitude
for each night is given in Table~1.

\section{Discussion}
\label{sec-discussion}
During observations made between 241 and 254~days post--outburst, we
detect aperiodic V-band variability in RS~Oph, with amplitudes ranging
from $\sim0.1-0.3$~mag, constituting `strong flickering' (Sokoloski et
al. 2001).  Observations made by \citet{Zam06} on day 117 of the 2006
outburst show no variability with amplitude above 0.03~mag.  In dwarf
novae, optical flickering is attributed to two sources: the turbulent
inner regions of the disc and the bright spot, where the stream of
matter from the Roche lobe-filling donor star impacts the outer edge
of the accretion disc, with inhomogeneities in the flow thought to
result in flickering \citep[e.g.][]{War95,Ken86}.  The physical
mechanism that causes flickering in symbiotics is not well understood,
but is believed to originate from accretion onto a white dwarf
\citep{Zam98}.  Adopting this assumption, these observations are
consistent with re-establishment of accretion between 117 and 241~days
after the onset of the 2006 outburst.  This is the earliest reported
detection of flickering subsequent to an outburst in RS~Oph.

\subsection{Mass transfer rate}
\label{sec-masstransfer}
Mass transfer from the secondary component is generally attributed to
one of two mechanisms: either Roche lobe overflow (RLOF) onto an
accretion disc; or through direct accretion of matter from the red
giant wind onto the white dwarf.  Assuming the flickering we observe
originates from a re-established accretion disc, we can place a
constraint on the mass transfer rate.  \citet{Sok03a} relate the time
taken to re-establish the disc (the viscous time-scale, $t_{\rm visc}$)
to the inner radius of the disc ($R_{\rm I}$).  This radius can be
further related to the rate of mass transfer through the disc (which
in this case we assume to equate to the white dwarf accretion rate,
$\dot{M}_{\rm acc}$) and the dynamical time-scale ($t_{\rm dyn}$), which
is approximately the time-scale of flickering.  Rearranging these
equations sourced from \citet{Fra92}, we find:

\begin{eqnarray*}
\dot{M}_{\rm acc} \sim 800 \left(\alpha \right)^{-8/3} \left(t_{\rm
visc}\right)^{-10/3} \left(t_{\rm dyn}\right)^{25/9}
\left(\frac{M_{\rm WD}}{M_{\sun}}\right)^{20/9}
\end{eqnarray*}

\noindent where $\dot{M}_{\rm acc}$ is in units of $M_{\sun} \rm
yr^{-1}$, and $\alpha$ depends on the state (high or low) of the disc,
with $\alpha = 0.03$ in the low state \citep{War95}, which we assume
in this case.  We take $M_{\rm WD}$ to be $1.35 M_{\sun}$
\citep{Hac00}.  As flickering recommenced between days 117 and 241, we
have a range of $1.01 \times 10^{7} \leq t_{\rm visc} \leq 2.08 \times
10^{7}$~s.  The shortest time-scale on which we see flickering is
$t_{\rm dyn} \approx 600$~s.  Thus for a low state we obtain an upper
limit of $\dot{M}_{\rm acc} \leq 4.1 \times 10^{-9}$ and a lower limit
of $\dot{M}_{\rm acc} \geq 3.7 \times 10^{-10} M_{\sun} \rm yr^{-1}$.

\subsection{Mass transfer mechanism}
\label{sec-nagaemodels}
In order to put this into context in terms of the mass transfer
mechanism operating in the system, we now consider these values
relative to mass transfer rates expected for accretion direct from the
red giant wind and via RLOF.  A mass accretion ratio, $f$, defined as
the ratio of the mass accreting onto the primary $\dot{M}_{\rm acc}$, to
the mass-loss rate from the donor companion $\dot{M}_{\rm giant}$, has
been calculated by \citet{Nag04}.  They quote $f~\le~1~\%$ in a
typical wind case, increasing to $f~\sim~10~\%$ for RLOF.  Studies of
the symbiotic star EG~And by \citet{Vog91} yield a mass-loss rate from
the red giant of $10^{-8} M_{\sun} \rm yr^{-1}$.  Since EG~And has a
number of similar parameters to RS~Oph (M2 red giant secondary,
483~day orbital period \citep{Fek00} cf. $\approx 460$~days in RS~Oph
\citep{Dob94}, similar absolute magnitude (Sokoloski et al. 2001)), we
adopt $\dot{M}_{\rm giant} \sim 10^{-8} M_{\sun} \rm yr^{-1}$ for
RS~Oph.  Applying the ratios from \citet{Nag04} to this mass loss rate
results in accretion rates of $\dot{M}_{\rm acc} \sim 10^{-9} M_{\sun}
\rm yr^{-1}$ for RLOF, and $\dot{M}_{\rm acc} \le 10^{-10} M_{\sun} \rm
yr^{-1}$ for direct wind accretion.  Thus our $\dot{M}_{\rm acc}$ limits
calculated in $\S$~\ref{sec-masstransfer} span the range required for
direct wind accretion and RLOF at the time accretion resumed.

\subsection{Outburst mechanism}
\label{outburst}

Since the outburst mechanism is dependent on the mass transfer rate,
 we now consider the implications of the rate determined for this
 early stage of resumed accretion.  \citet{Yar05} present a grid of
 outburst characteristics compiled from models of thermonuclear
 runaway in novae.  These data predict that for a system with a mass
 transfer rate of $10^{-9}$ to $10^{-10}$~$M_{\sun} \rm yr^{-1}$ onto
 a hot 1.4 $M_{\sun}$ white dwarf, we should expect an outburst
 recurrence period ranging from 200 to over 1000~yr, whereas the time
 elapsed between observed outbursts in RS~Oph averages $\sim$20~yr.
 Indeed, translating this model to a slightly lower white dwarf mass
 more appropriate for RS~Oph (i.e. $1.35 M_{\sun}$ from \citet{Hac00})
 produces a further increase in the outburst recurrence interval since
 the accreted mass required to trigger thermonuclear runaway is higher
 for a lower mass white dwarf.  To allow for discrepancies in the
 white dwarf mass, basing these calculations on the value of $1.2
 M_{\sun}$ determined by \citet{Sta96} results in a lower $\dot{M}_{\rm
 acc}$, lengthening the recurrence interval still further.  From
 \citet{Yar05}, a recurrence period of $\sim$20~yr is achievable only
 if we have 100\% accretion efficiency, i.e. $\dot{M}_{\rm acc} =
 \dot{M}_{\rm giant} = 10^{-8} M_{\sun} \rm yr^{-1}$, which far exceeds
 the findings of e.g. \citet{Nag04} ($\S$~\ref{sec-nagaemodels}).
 While our upper limit on the accretion rate approaches $10^{-8}
 M_{\sun} \rm yr^{-1}$, the non-linear relation of the \citet{Yar05}
 model means that the recurrence period remains several times longer
 than 20~yr for $\dot{M}_{\rm acc} \sim 4 \times 10^{-9} M_{\sun} \rm
 yr^{-1}$, and indeed a factor of two greater than the longest
 interval between observed outbursts in this system.


The accretion luminosity of the system would be most accurately
measured at UV wavelengths. While observations were made in the UV
with $Swift$, none exists prior to day 25 (Goad \& Beardmore, private
communication).  From this point on the UV tracks the behaviour of the
supersoft X-ray emission attributed to fusion on the white dwarf
surface \citep{Hac07}.  The 1985 observations came at a similar point
post--outburst.  Hence between outbursts we need to estimate accretion
rates by less direct methods.  Standard accretion theory predicts that
disc luminosity is proportional to the mass transfer rate
\citep{Zam98}.  Thus the visual quiescent variation of 2.5~mag
reported by \citet{Opp93} implies a factor 10 variation in mass
transfer rate during quiescence.  As the visual magnitude during our
observations was at the lower end of the quiescent magnitude range
this implies that the inter-outburst accretion rate is typically
higher than we see here.




Such variations of mass transfer rate are plausible in either the RLOF
or wind accretion scenario; either on short time-scales due to erratic
or clumpy mass transfer, or over longer periods, perhaps increasing as
the disc becomes better established.  \citet{Hac00}, for example,
determine a much larger mass accretion rate of $\dot{M}=1.2 \times
10^{-7} M_{\sun} \rm yr^{-1}$ for RS~Oph between the outbursts in 1967
and 1985, and brightness variations of up to 3~mag have been observed
during periods of quiescence \citep{Ros87}.  Furthermore, recurrence
intervals in this object vary from 9 to 35 yr.  Orbital eccentricity
may have a particularly marked effect on the rate of mass transferred
by direct wind accretion, as the white dwarf trajectory would trace a
route through varying densities of the red giant wind.  Indeed, the
eccentricity in the system is completely unconstrained; \citet{Dob96}
quote $e = 0.25 \pm 0.70$ when modeled using the giant component and
$e=0.40 \pm 1.40$ using the white dwarf.  Another factor not accounted
for in the models that could potentially cause inconsistencies in the
nova recurrence interval is that of residual heating of the white
dwarf following an outburst, lowering the accreted mass required to
trigger a subsequent outburst.  Further work is needed to fully verify
the outburst mechanism in this and similar systems.

\section{Conclusions}
\label{sec-conclusions}

\begin{enumerate}
\item Statistically significant flickering is detected in RS~Oph on days
    241 to 254 of the 2006 outburst, consistent with the re-establishment of
    accretion between days 117 and 241 after outburst.

\item Over the 2 week period of observations, the mean V magnitude
    decreases by $\sim 0.5$~mag from 11.4~to~11.9~mag.

\item Calculated limits on the white dwarf accretion rate of $4\times
10^{-10} \ltsimeq \dot{M}_{\rm acc} \ltsimeq 4 \times 10^{-9} M_{\sun}
\rm yr^{-1}$ span the range required for both direct wind accretion
and RLOF mechanisms.  We therefore find no conclusive evidence
favouring one accretion mechanism over the other in RS~Oph.




\item Current models are not sufficiently complete to confidently
determine the accretion and outburst mechanisms in RS~Oph.

\end{enumerate}   

\section*{Acknowledgements}

\noindent We thank Dave Kilkenny and Lisa Crause for their invaluable assistance, and the SAAO TAC for generous allocation of telescope time.  HLW acknowledges studentship support from the University of Central Lancashire.  This paper uses observations made at the South African Astronomical Observatory (SAAO).  IRAF is distributed by the National Optical Astronomy Observatories, which are operated by the Association of Universities for Research in Astronomy, Inc., under cooperative agreement with the National Science Foundation.

\bsp

\label{lastpage}
\end{document}